\documentclass{PoS}
\usepackage{amsmath}
\usepackage{multicol}
\title{Mobility edge and locality of the overlap-Dirac operator with
       and without dynamical overlap fermions}

\ShortTitle{Mobility edge and locality of the overlap-Dirac operator
       with and without dynamical $\cdots$}

\author{JLQCD Collaboration:
\speaker{N.~Yamada}\,$^{a,b}$\thanks{E-mail: norikazu.yamada@kek.jp},~
S.~Aoki$^{\,c,d}$, H.~Fukaya$^e$,
S.~Hashimoto$^{a,b}$, K-I.~Ishikawa$^f$, K.~Kanaya$^c$,
T.~Kaneko$^{a,b}$, H.~Matsufuru$^a$,~M.~Okamoto$^a$,  T.~Onogi$^g$\\
\llap{$^a$}High Energy Accelerator Research Organization (KEK), Tsukuba
           305-0801,Japan\\
\llap{$^b$}School of High Energy Accelerator Science, The Graduate
           University for Advanced Studies (Sokendai), Tsukuba
           305-0801, Japan\\
\llap{$^c$}Graduate School of Pure and Applied Sciences, University of
           Tsukuba, Tsukuba 305-8571, Japan\\
\llap{$^d$}Riken BNL Research Center, Brookhaven National Laboratory,
           Upton, New York 11973, USA\\
\llap{$^e$}Theoretical Physics Laboratory, RIKEN, Wako 351-0198, Japan\\
\llap{$^f$}Department of Physics, Hiroshima University,
           Higashi-Hiroshima 739-8526, Japan\\
\llap{$^g$}Yukawa Institute for Theoretical Physics, Kyoto University,
           Kyoto 606-8502, Japan
}
\abstract{
\vspace*{-100ex}
\begin{flushright}
 HUPD-0608, KEK-CP-181, RIKEN-TH-79
\end{flushright}
\vspace*{100ex}
We perform a systematic study of low-lying eigenmodes of $H_w$ with
various gauge actions to find the optimal choice for dynamical overlap
fermion simulations, with which one may achieve lower numerical cost for
HMC and better locality property of the overlap kernel.
For this purpose, our study is made with emphasis on the distribution of
low-lying eigenvalues and the mobility edge with and without dynamical
overlap fermions.
}

\FullConference{XXIV International Symposium on Lattice Field Theory\\
		 July 23-28 2006\\
		 Tucson Arizona, US}

\begin{document}

\section{Introduction}

The JLQCD collaboration has started a large scale simulation of
dynamical overlap fermions~\cite{Fukaya:lattice2006}, aiming at studying
the QCD dynamics in the presence of very light quarks.
For the overlap fermion formulation, large computational costs and
non-locality of the overlap-Dirac operator have been practical and
theoretical concerns.
The computational cost is closely related to the near-zero mode density
of the Hermitian Wilson-Dirac operator,
\begin{eqnarray}
     H_{\rm w}(-m_0)
 &=& \gamma_5 D_{\rm w}(-m_0)
  =  \gamma_5 \left(-m_0+D_{\rm w}(0)\right)\ \ \ \
     (m_0= 1+s),
\end{eqnarray}
where $D_{\rm w}(0)$ is the standard Wilson-Dirac operator for the
massless Wilson fermion.
According to Ref.~\cite{Golterman:2003qe}, locality of the overlap-Dirac
operator can be studied by examining the locality of eigenvectors of
$H_{\rm w}$, defined by
$H_{\rm w}\phi_i(x)=\lambda_{\rm w,\it i}\phi_i(x)$. 
Since both the near-zero mode density and the corresponding
eigenvectors are known to depend on the gauge action and $\beta$, in
this work we explore these two properties for several gauge actions.

We consider three gauge actions, the standard plaquette (Plq), Iwasaki (RG),
and a modified gauge action motivated by the admissibility condition
(Adm).
The Adm action~\cite{Luscher:1998du,Fukaya:2005cw,Bietenholz:2005rd} is
given by
\begin{eqnarray}
 S_{\rm Adm}=\left\{\begin{array}{cl}
  \beta\,\sum_{x,\mu,\nu}\displaystyle
   \frac{1-{\rm Re}{\rm Tr}P_{\mu,\nu}(x)/3}
        {1-(1-{\rm Re}{\rm Tr}P_{\mu,\nu}/3)/\epsilon},
  &~~~\mbox{when }~~ 1-{\rm Re}{\rm Tr}P_{\mu,\nu}/3<\epsilon\\
   \infty &~~~\mbox{otherwise}
               \end{array}\right.,
\end{eqnarray}
where $\epsilon$ is a parameter to control the possible maximum value of
${\rm Re}{\rm Tr}P_{\mu,\nu}(x)/3$.
With $\epsilon<1/(6(2+\sqrt{2}))$, the locality of the overlap-Dirac
operator is guaranteed~\cite{Neuberger:1999pz}.
This action reduces to the standard plaquette action when
$1/\epsilon\rightarrow 0$.
In addition, for each of these gauge actions we introduce two-flavors of
extra-Wilson fermions $\psi$ and ghosts $\chi$~\cite{Izubuchi:2002pq} as
\begin{eqnarray}
   S_{\rm ext}
 = \sum_x \bar\psi(x)D_{\rm w}(-m_0)\psi(x)
 + \sum_x \chi^\dag(x)\left(D_{\rm w}(-m_0) + i\mu\gamma_5\tau^3
                      \right)\chi(x),
\end{eqnarray}
where $\tau^3$ acts on the flavor index.
Since integrating out these extra fields results in
$\det|H^2_{\rm w}(-m_0)$ $/(H^2_{\rm w}(-m_0) + \mu^2)|$, the appearance
of the modes with $|\lambda_{\rm w,\it i}|<\mu$ are suppressed with this
weight.
$\mu=0$ corresponds to the standard quenched approximation.
The same value of $m_0$ is taken for both the overlap kernel and extra
fields.

The lattice size is fixed to $16^3\times 32$.
For the Adm action, $1/\epsilon=2/3$ is used throughout this study.
Three values of $\mu$ (0.0, 0.2, and 0.4) are examined, and more than
10,000 trajectories are accumulated for each gauge action as shown in
Tab.~\ref{tab:sim para}.
The lattice spacing is set by $r_0$=0.49 fm, and is about 0.125 fm for
all lattices, unless otherwise stated.

\section{Spectral density}

\begin{figure}[h]
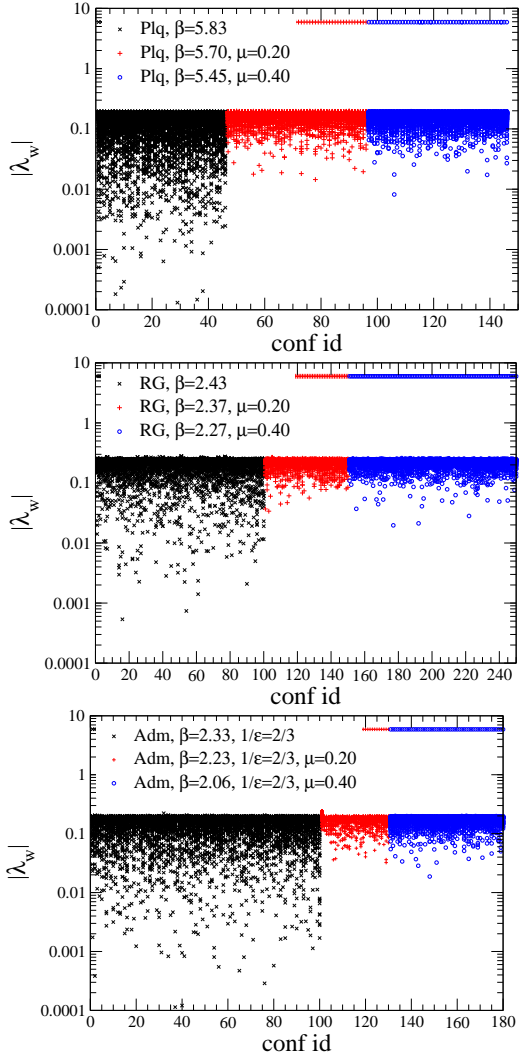

 \vspace{-1ex}
 \centering
   \includegraphics*[width=0.45 \textwidth,clip=true]
   {figs/abs_qplq.eps}\\
   \includegraphics*[width=0.45 \textwidth,clip=true]
   {figs/abs_qrg.eps}\\
   \includegraphics*[width=0.45 \textwidth,clip=true]
   {figs/abs_qadm.eps}
 \vspace{-2ex}\\
 \caption{Distribution of near-zero modes for three different gauge
 actions, Plq, RG and Adm from top to bottom, and with three values of
 $\mu$=0.0, 0.2, 0.4 from left to right.
 }
 \label{fig:eigen dist}
 \vspace{-22ex}
\end{figure}

\begin{figure}[h]
 \centering
   \includegraphics*[width=0.45 \textwidth,clip=true]
   {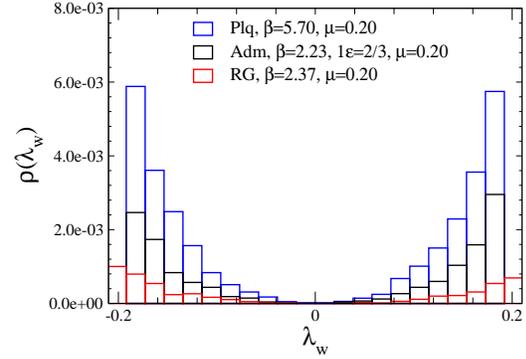}
 \vspace{-2ex}\\
 \caption{Spectral densities for three quenched gauge actions, Plq
 (blue), Adm (black) and RG (red) with $\mu=0.2$.}
 \label{fig:eigen hist}
 \vspace{-0ex}
\end{figure}

Three panels in Fig.~\ref{fig:eigen dist} show the distributions of the
near-zero modes for three gauge actions, where all eigenvalues are
plotted in the lattice unit.
Different symbols represent different values of $\mu$.
We find that the extra fermions and ghosts suppress the low-lying modes
as expected.
The suppression is most effective with $\mu=0.2$.
Figure~\ref{fig:eigen hist} shows a comparison of the spectral
density $\rho(\lambda_{\rm w})$ for the three gauge action with
$\mu=0.2$.
We observe that the RG action yields the smallest near-zero mode
density.\ 
From these observations made in the quenched approximation, we decide
to employ the Iwasaki RG gauge action with $\mu=0.2$ in the dynamical
overlap simulations.
\twocolumn
\begin{table}[h]
 \vspace{-1ex}
 \centering
 \begin{tabular}{c|cccc}
  action & $\beta$ & $\mu$ & $1/\epsilon$ & \# of trj.\\
\hline
  Plq & 5.83 & 0.0 & & 20,000\\
      & 5.70 & 0.2 & & 11,600\\
      & 5.45 & 0.4 & & 11,600\\
\hline
  RG  & 2.43 & 0.0 & & 20,000\\
      & 2.37 & 0.2 & & 21,600\\
      & 2.27 & 0.4 & & 20,000\\
\hline
  Adm & 2.33 & 0.0 & 2/3 &  20,000\\
      & 2.23 & 0.2 & 2/3 &  20,000\\
      & 2.06 & 0.4 & 2/3 &  14,800\\
 \end{tabular}
 \caption{Simulation parameters.}
 \vspace{-0ex}
 \label{tab:sim para}
\end{table}
\begin{figure}[h]
 \centering
 \begin{tabular}{c}
   \includegraphics*[width=0.45 \textwidth,clip=true]
   {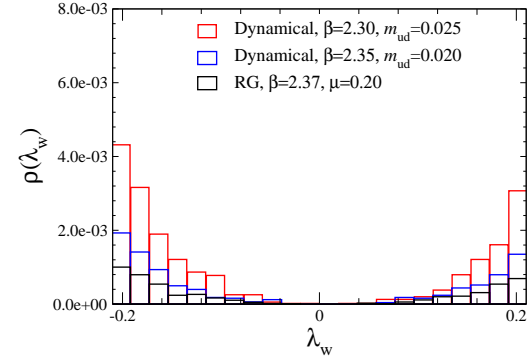}
 \end{tabular}
 \vspace{-2ex}\\
 \caption{Comparison of the near-zero mode densities on the quenched
 (black), slightly fine (blue) and coarse (red) dynamical
 configurations.}
 \label{fig:comp quench dynamical}
 \vspace{-2ex}
\end{figure}
\vspace*{17ex}
\ \ \ \

We made the similar study on dynamical configurations, and compare the
results with that on the quenched results with $\mu$=0.2 in
Fig.~\ref{fig:comp quench dynamical}.
The results from two dynamical ensembles with $\beta$ = 2.30 and 2.35
are shown, their lattice spacings are $a$=0.12 fm and 0.11 fm,
respectively, and the sea quark mass is about $m_s/4$ in both cases.
While the density increases for the dynamical configurations, it is
clear that the near-zero mode suppression works well even after
incorporating the overlap sea quarks.
More details on our dynamical configurations are found in
Ref.~\cite{Fukaya:lattice2006}.

\section{Mobility edge}

The overlap-Dirac operator is proved to be exponentially local, if there
is no low-lying mode 
\onecolumn
\hspace*{-4.8ex}
below some threshold~\cite{Hernandez:1998et}.
In practice, the
density of the near-zero mode is non-zero, unless we introduce the
determinant factor such as $\det H_{\rm w}^2$.
Even in this case, the overlap-Dirac operator is exponentially local if
the near-zero modes themselves are exponentially
localized~\cite{Golterman:2003qe}.
Golterman, Shamir, and Svetitsky argued that the magnitude of the
overlap operator behaves as
\begin{eqnarray}
     |D_{\rm ov}(x,y)|
\sim \bar\lambda\rho(\bar \lambda)
     \exp\left(-\frac{|x-y|}{2 l_l(\bar\lambda)}\right)
   + O(1)\cdot\exp(-\lambda_c|x-y|),
   \label{eq:conjecture}
\end{eqnarray}
where $\bar\lambda$ denotes a near-maximum eigenvalue of low-lying
localized modes (somewhat ambiguous), $l_l(\bar\lambda)$ a localization
length at $\bar\lambda$.
The parameter $\lambda_c$ stands for the mobility edge, which separates
localized and extended modes.
The first term is derived from the dependence of $\rho(\lambda)$ and
$l_l(\lambda)$ on $\lambda$, while the second term is a
conjecture motivated by numerical experiences~\cite{Golterman:2003qe}.
In most cases, it is known that the second term dominates the first.

\begin{figure}[b]
 \centering
 \begin{tabular}{c}
   \includegraphics*[width=0.6 \textwidth,clip=true]
   {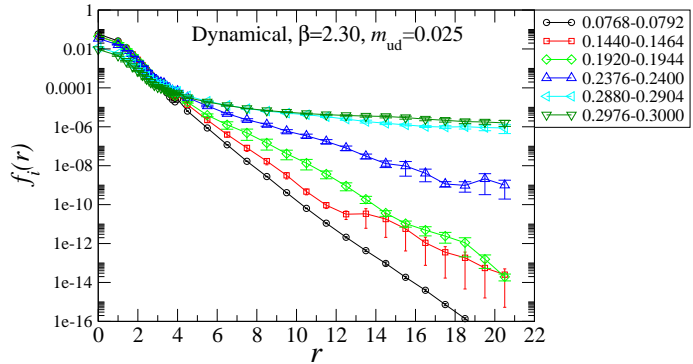}
 \end{tabular}
 \vspace{-2ex}\\
 \caption{Example of $f_i(r)$.
   Different symbols represent different bin in $\lambda_{\rm w}$, as
 denoted in the plots.
 Data for the dynamical configuration at $\beta$=2.30, $m_{\rm
 sea}$=0.025.}
 \label{fig:fr}
 \vspace{-2ex}
\end{figure}

Mobility edge is determined from the eigenvectors as follows.
We first define $\rho_i(x)$ and $f_i(r)$ as
\begin{eqnarray}
    \rho_i(x)
&=& \phi_i^\dag(x)\phi_i(x),~~~
    \rho_i(x_0)=\max_x\{\rho_i(x)\},\\
    f_i(r)
&=& \{\bar\rho_i(x)\,\big|\,r=|x-x_0|\},
\end{eqnarray}
where $\bar\rho_i(x)$ is an average of $\rho_i(x)$ over the lattice
points which have the same distance $r$ from $x_0$.
The eigenvalues $|\lambda_{\rm w,i}|$ are binned with a certain bin
size, and $f_i(r)$ is averaged over the modes within a bin.
The localization length at each bin, $l_l(\lambda_{\rm w,i})$, is then
obtained at large $r$ by fitting to
\begin{eqnarray}
 f_i(r)=\exp\left(-\frac{r}{l_l(\lambda_{\rm w,i})}\right).
 \label{eq:fit func}
\end{eqnarray}
Mobility edge, $\lambda_c$, is defined by $\lambda_{\rm w,i}$ at which
$l_l(\lambda_{\rm w,i})$ diverges.
Figure~\ref{fig:fr} shows an example of $f_i(r)$ obtained on a single
configuration.
We can clearly see that the decay rate becomes smaller for
larger eigenvalues.

\begin{figure}[t]
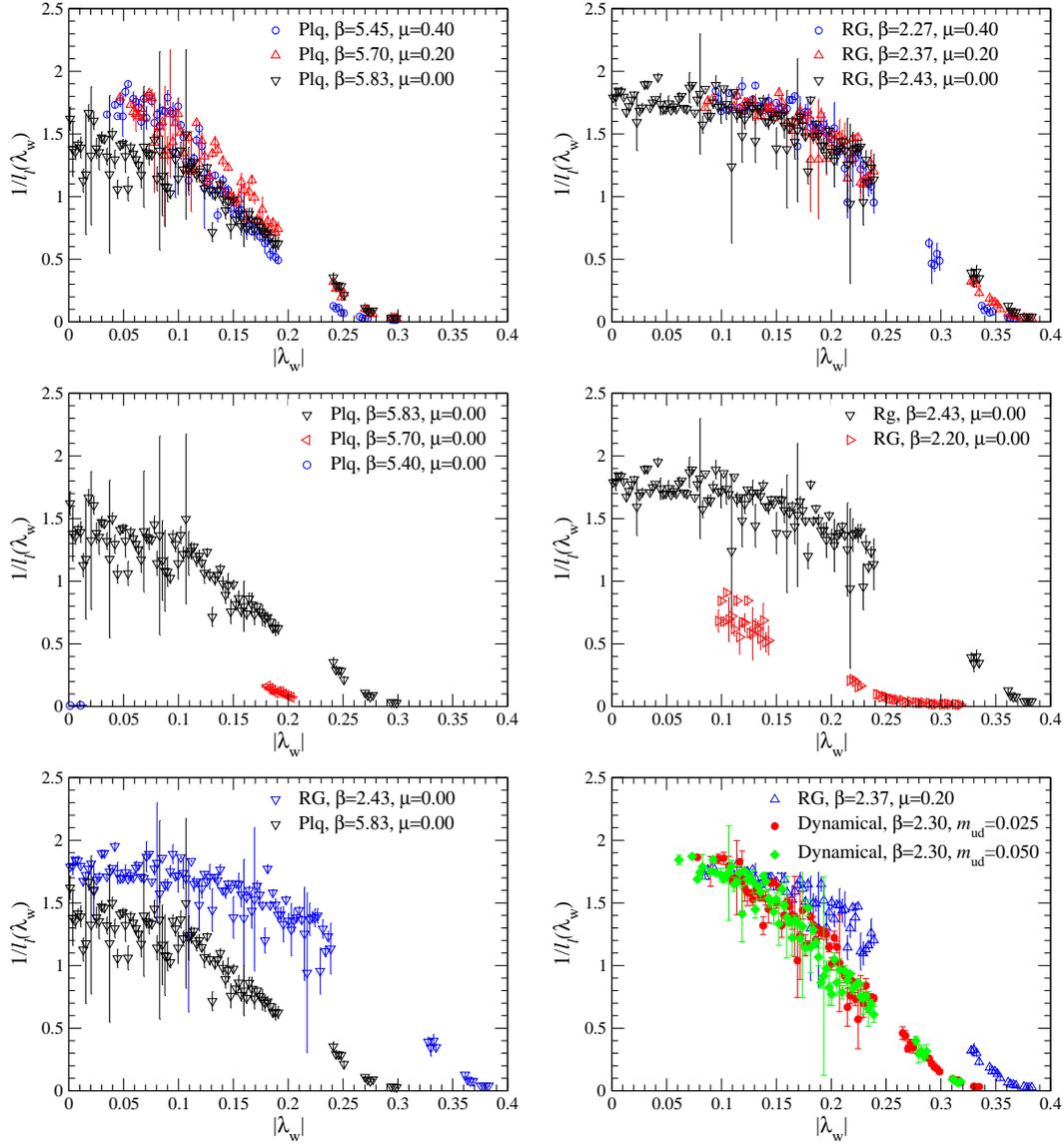

 \centering
 \begin{tabular}{cc}
   \includegraphics*[width=0.45 \textwidth,clip=true]
   {figs/mobedge.qplq.eps}&
   \includegraphics*[width=0.45 \textwidth,clip=true]
   {figs/mobedge.qrg.eps}\\
   \includegraphics*[width=0.45 \textwidth,clip=true]
   {figs/mobedge.qplq_beta.eps}&
   \includegraphics*[width=0.45 \textwidth,clip=true]
   {figs/mobedge.qrg_beta.eps}\\
   \includegraphics*[width=0.45 \textwidth,clip=true]
   {figs/mobedge.quench.eps}&
   \includegraphics*[width=0.45 \textwidth,clip=true]
   {figs/mobedge.drg_comp.eps}\\
 \end{tabular}
 \vspace{-2ex}\\
 \caption{Inverse localization length $1/l_l(\lambda_{\rm w})$ as a
 function of $|\lambda_{\rm w}|$.}
 \label{fig:l_l}
 \vspace{-1ex}
\end{figure}
The determination of $\lambda_c$ is performed with a fixed value of
$m_0$=1.6.
In addition to the ensembles used in the study of spectral density, the
determination is performed on three coarse lattices, which are generated
with the Plq action with $\beta=$5.70 and 5.40 and the RG action with
$\beta$=2.43, to see the dependence on lattice spacing.
$l_l(\lambda_{\rm w,\it i})$ extracted from the fit using
eq.~(\ref{eq:fit func}) is plotted as a function of
$\lambda_{\rm w,\it i}$ in Fig.~\ref{fig:l_l} for various ensembles. 
In the two plots at the top of Fig.~\ref{fig:l_l}, the results at the
same lattice spacing, $a$=0.125 fm, but with different $\mu$ are plotted
for the Plq (left) and the RG (right).
$\lambda_c$ is found to be 0.24-0.27 for the plaquette action and
0.34-0.36 for the RG action in the lattice unit.
Dependence of $\lambda_c$ on $\mu$ turns out to be weak in the range of
$0.0<\mu<0.4$.

From the two plots in the middle showing the results with $\mu=0$ for
two or three different $\beta$, we find that $\lambda_c$ in the lattice
unit decreases as $\beta$ decreases.
The data at $\beta=5.40$ for the plaquette action (circle in the
left-middle panel) shows $1/l_l(0)\sim 0$.
Ref.~\cite{Golterman:2003qe} suggests to use ``$1/l_l(0)=0$'' as the
definition of the Aoki phase.

In the left-bottom, the results with the plaquette and RG actions are
compared at the same lattice spacing.
Clearly the RG action gives lager $\lambda_c$ than that of the Plq
action.
Finally, the right-bottom figure shows a comparison of the results from
quenched and dynamical runs with two different sea quark masses.
Here we do not see any significant difference between the quenched and
dynamical runs nor clear dependence on sea quark mass within dynamical
runs.
$\lambda_c$ turns out to be 500--600 MeV in our dynamical lattices.

We also measure $1/l_{\rm ov}$ from the $r$ dependence of $h(r)$
defined by
\begin{eqnarray}
 h(r)=\max_x\left\{
      ||D_{\rm ov}(x,y)\delta(y-x_0)||~ \Big|~
      r=\sum_\mu|x_\mu-{x_0}_\mu|
          \right\}.
\end{eqnarray}
 \begin{figure}[t]
  \centering
  \includegraphics*[width=0.45 \textwidth,clip=true]
  {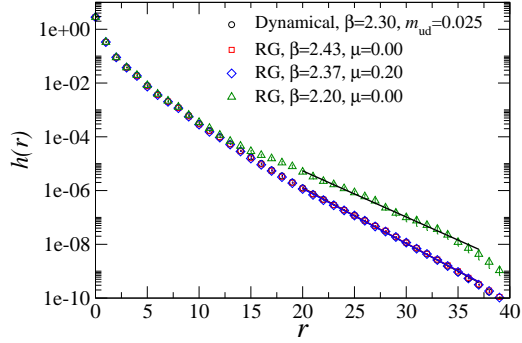}
  \caption{$r$ dependence of $h(r)$ for four different ensembles.}
  \label{fig:lov}
 \end{figure}
$h(r)$ is measured on the same ensembles, and some of the results
obtained with the RG action are plotted in Fig.~\ref{fig:lov}, where
numerical data are in the lattice unit.
We see that all results except for the one with $\beta$=2.20 coincide
with each other.
$\beta$=2.20 corresponds to $a\approx 0.2$ fm, while others are
$a$=0.120-0.125 fm.
By fitting this to the same form as eq.~(\ref{eq:fit func}), we obtain
$1/l_{\rm ov}$.
For our dynamical lattice at $\beta$=2.30, $1/l_{\rm ov}$ is estimated
to be about 800 MeV.

\begin{figure}[b]
 \centering
   \includegraphics*[width=0.45 \textwidth,clip=true]
   {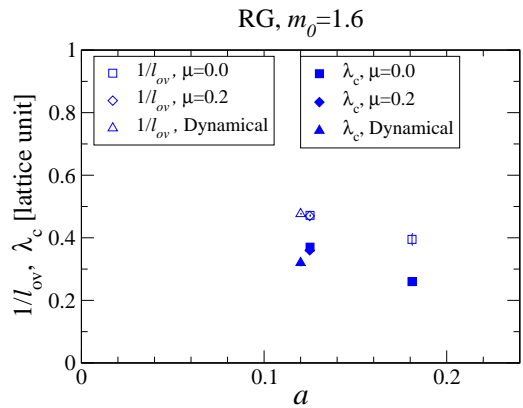}
 \caption{$a$ dependence of $1/l_{\rm ov}$ and $\lambda_c$.}
 \label{fig:lov lambda_c}
\end{figure}
In Fig.~\ref{fig:lov lambda_c}, $\lambda_c$ and $1/l_{\rm ov}$ obtained
with the RG action are compared.
While both $1/l_{\rm ov}$ and $\lambda_c$ show the similar dependence on
lattice spacing, the difference is sizable.
Within the range of the lattice spacing we have studied it turns out
that $1/l_{\rm ov}>\lambda_c$.
Although $\lambda_c\sim$500-600 MeV in our dynamical lattice is not much
larger than $\Lambda_{\rm QCD}$, $1/l_{\rm ov}\sim$ 800 MeV is probably 
acceptable for simulations of QCD.
In Ref.~\cite{Draper:2005mh}, $1/l_{\rm ov}$ has been studied at several
different lattice spacings in the range of $a\sim$0.13--0.20 fm within
the quenched approximation.
They reported acceptably large values for $1/l_{\rm ov}$ at all lattices,
which appears to be consistent with our observation.

\section{Summary}

In this work, near-zero mode density for various gauge actions are
examined from the viewpoint of cost reduction in dynamical overlap
simulation, and the RG Iwasaki action with extra fermions and ghosts
exhibits smallest near-zero mode density among the actions studied.
We also determine the mobility edge and the localization range of
the overlap-Dirac operator.
While $\lambda_c$ and $1/l_{\rm ov}$ seem to agree with each other
qualitatively, the difference is sizable.
$1/l_{\rm ov}$ tells us the locality of an given overlap-Dirac operator
and is presumably all we need to know from the practical point of view,
but the precise relationship between $1/l_{\rm ov}$ and $\lambda_c$
should be understood.

\section*{Acknowledgment}

Numerical simulations are performed on Hitachi SR11000 and
IBM System Blue Gene Solution
at High Energy Accelerator Research Organization (KEK) under a
support of its Large Scale Simulation Program (No.~06-13).
This work is supported in part by the Grant-in-Aid of the Ministry of
Education (Nos. 13135204, 13135213, 15540251, 16740147, 16740156,
17340066, 17740171, 18034011, 18340075, 18740167.)


\end{document}